# Deployment Opportunities for DPS-QKD in the Co-Existence Regime of Lit GPON / NG-PON2 Access Networks

**Nemanja Vokić, Dinka Milovančev, Bernhard Schrenk, Michael Hentschel, and Hannes Hübel**
*AIT Austrian Institute of Technology, Center for Digital Safety&Security / Security & Communication Technologies, 1210 Vienna, Austria.*
*Author e-mail address: bernhard.schrenk@ait.ac.at*

We demonstrate cost-effective QKD integration for GPON and NG-PON2. Operation at $5.1\times10^{-7}$ secure bits/pulse and a QBER of 3.28% is accomplished for a 13.5-km reach, 2:16-split PON, with 0.52% co-existence penalty for 19 classical channels.

## 1. Introduction

Quantum key distribution (QKD) beneficially exploits the laws of quantum mechanics to securely provide a secret key between two communication parties. A variety of QKD protocols has been demonstrated [1], however, recent QKD developments are still being subject to high deployment cost. Furthermore, the practical network integration without resorting to dedicated dark fibers is of high interest. Raman scattering, which is characterized by its wide spectral tails, has been identified as the most critical impairment when co-existence with classical channels is to be accomplished [2]. For the particular case of passive optical access networks (PON), a segment where cost efficiency is paramount for practical deployment, the high splitting loss of the optical distribution network (ODN) cannot be simply bypassed through insertion of waveband filters. Furthermore, the wavelength plan of PON standards leads to Raman crosstalk from the O- to the L- band, thus rendering the QKD channel allocation as challenging.

In this work, we focus on a techno-economic efficiently differential phase shift (DPS) QKD integration scheme for PONs. Through a technologically lean end-user QKD sub-system based on a single laser and consolidation of more specific quantum-optics, such as single-photon detectors, at a centralized location amenable to cost-sharing, a favorable asymmetry in complexity is obtained between head- and tail-end of the link. We will show that DPS-QKD operation in a lit (i.e. dark-fiber free) network can be accomplished in presence of classical signals of the GPON and NG-PON2 access standards. Secure-key generation at a rate of $5.1\times10^{-7}$ bits/pulse and a QBER of 3.28% is obtained over a 13.5-km reach, 2:16-split PON in co-existence with carrier-grade classical channels in the O/S/C/L-bands.

## 2. Integration of Differential Phase Shift QKD in Lit Passive Optical Networks

A typical PON setting with integrated QKD channel is illustrated in Fig. 1. Classical channels at the central office (CO) and the optical network units (ONU) induce Raman scattering at the fibers, which determines the integration of the sensitive quantum channel. The classical upstream $C_{US}$ features fewer lanes than the downstream $C_{DS}$, which is upgraded by a WDM overlay for wireless fronthauling. Raman scattering can be reduced by either a directional split between quantum and classical signals, or by means of narrowband optical filtering. The implementation of the quantum channel $Q_{US}$ is more advantageous in upstream direction: Downstream-induced Raman noise $\delta_F$ at the dominant feeder length is avoided by a dual-feeder ODN containing a high-directivity 2:$N$ splitter. The weaker noise $\delta_D$, which arises at the shorter drop fibers, will be attenuated by to splitting loss. Moreover, since the classical upstream $C_{US}$ follows time division multiple access (TDMA), the overall noise of all $N$ drop fibers will correspond to a single continuous-mode signal. It passes the PON over the entire reach of feeder ($\upsilon_F$) and drop ($\upsilon_D$) fibers. Finally, the constant dark count rate $\Delta$ of the single-photon avalanche detector (SPAD) at the CO needs to be taken into consideration. It shall be further stressed that the high splitting ratios 1:$M\times N$ of the PON do not necessarily have to be implemented at the tree. Often, a first split 1:$M$ is realized at the consolidation point at the CO, which allows to bypass part of the splitting loss when multiplexing the QKD channel.

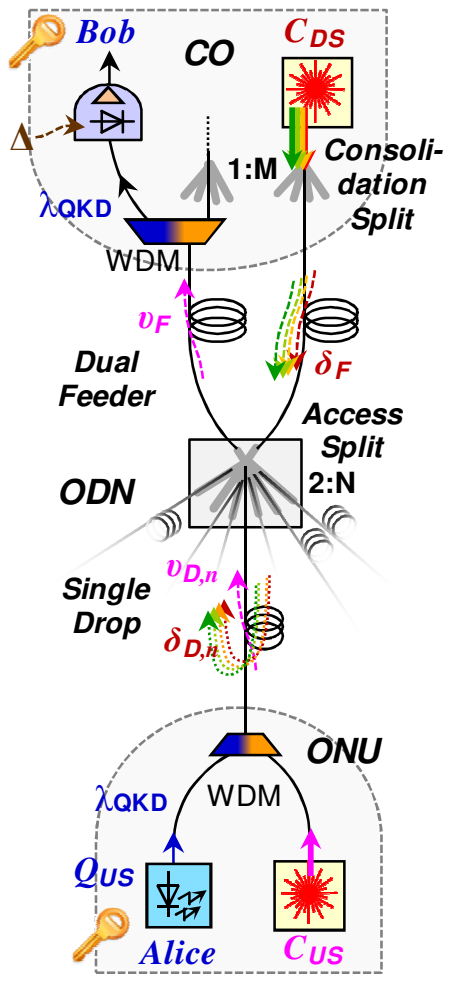

Fig. 1. PON integration of DPS-QKD.

## 3. Experimental Setup and Network Load

The experimental setup to evaluate DPS-QKD integration in the PON is presented in Fig. 2a. The DPS transmitter at the ONU is based on a directly chirp-modulated laser (DML), which performs binary optical phase modulation in conjunction with a Mach-Zehnder modulator (MZM) assisted pulse carver to further suppresses the symbol edges. In principle, both tasks could be performed by a single, integrated externally modulated laser (EML) device, as it has been recently demonstrated [3]. The DPS signal is then launched at a mean photon number of $\mu = 0.1$.

The DPS receiver at the CO is composed of an optical phase demodulator and a free-running SPAD with an efficiency of 10%. An asymmetric delay interferometer (DI) for a DPS symbol rate of 1 Gbaud was used for demodulation. Its extinction is optimized using a polarization controller (PC). Detection events are registered by a time-tagging module (TTM), which performs a real-time estimation of the raw-key rate and the QBER.

The ODN of the PON as the transmission channel consists of a feeder with lengths of 15.2 and 13.2 km in down- and upstream, respectively. After the 2:$N$ tree splitter, a 256-m drop fiber connects to the ONU. The average losses of these standard single-mode fibers (SMF) were 0.21 dB/km (at 1550 nm) and 0.39 dB/km (at 1310 nm).

The network was loaded according to the GPON and NG-PON2 standards (Fig. 2b). For GPON the quantum channel is placed at the C-band ($\lambda_{QKD}$ = 1550.12 nm), above the downstream $C_{DS}$ at 1489 nm and the upstream $C_{US}$ at 1310 nm. Optical waveband add/drop (A/D) filters clean the classical emission from far-reaching ASE tails. The GPON power levels were 2.2 and 0.3 dBm. These levels are realistic considering a first split stage at the CO site and the typical loss budget of deployed PONs of ~22 dB [4] below the class B+ budget. In the NG-PON2 case, the O-band is used for QKD integration ($\lambda_{QKD}$ = 1310.55 nm). Figure 2b shows the classical network load: four L-band downstream channels $C_{DS}$, 11 upper C-band fronthaul channels $C_{FH}$, and four lower C-band upstream channels $C_{US}$.

The QDK channel is demultiplexed at the CO from the classical upstream through two red/blue (R/B) waveband filters. A narrowband filter is cascaded with these broadband filters in order to define a narrow reception bandwidth and thus prevents excessive Raman noise. Three filter types have been chosen for this task: a 800-GHz LAN-WDM filter in the O-band (NG-PON2), a 100-GHz DWDM A/D filter in the C-band (GPON), and a spectrally narrow fiber Bragg grating (FBG) with a bandwidth of 14.6 GHz. The transfer curves of these filters are reported in Fig. 2c.

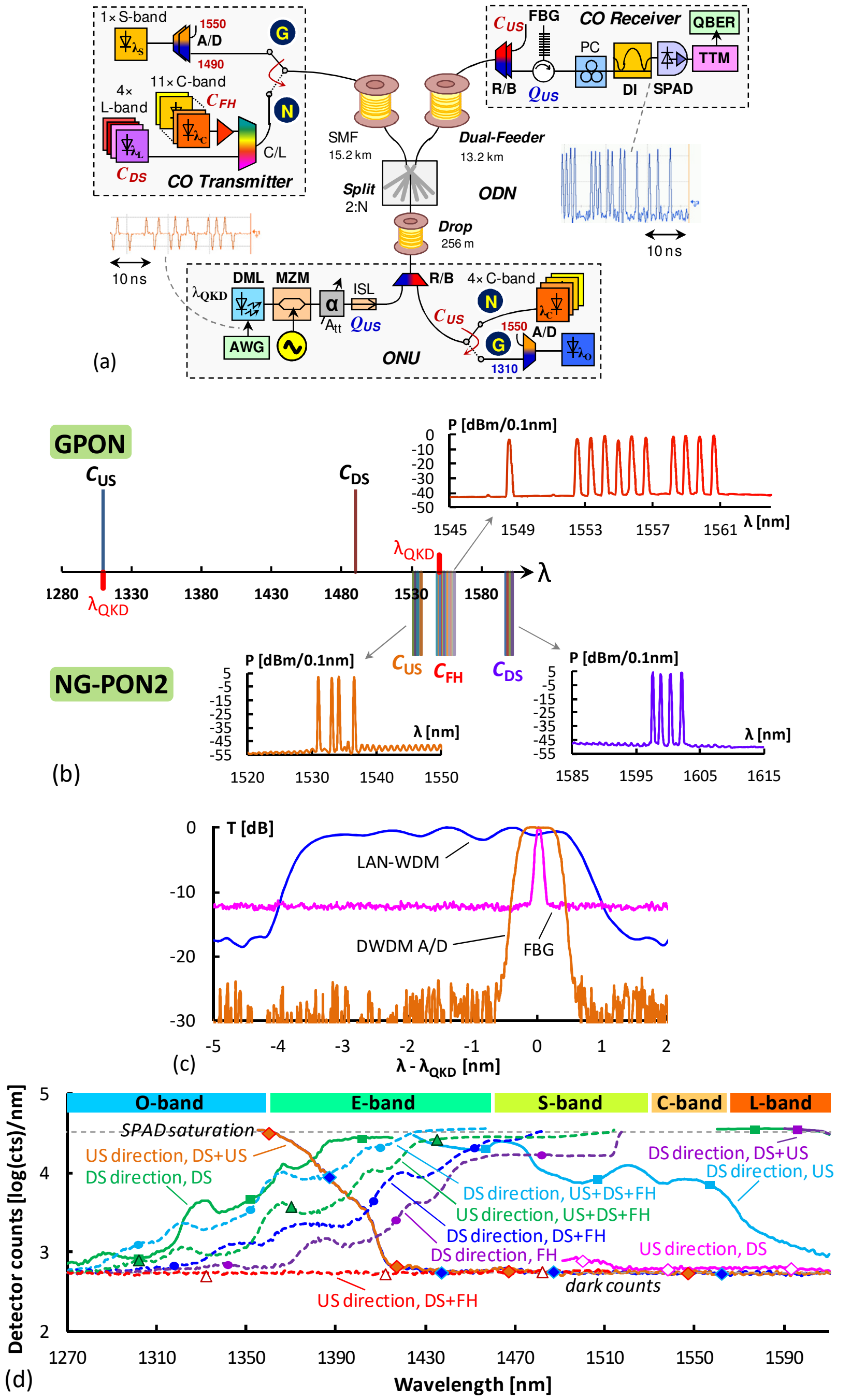

Fig. 2. (a) Experimental setup for GPON (G) and NG-PON2 (N) integrated QKD. (b) Optical spectrum for launched NG-PON2 channels. (c) Narrowband filters used at the DPS receiver. (d) Raman noise induced at a 2:16 split PON by GPON (solid) and NG-PON2 (dashed lines).

## 4. Noise Contribution due to Stimulated Raman Scattering and DPS QKD Performance

Figure 2d shows the Raman noise spectra that was acquired through a single-photon optical spectrum for a 2:16 split PON. Measurements are shown for GPON-induced noise (solid lines) in down- (■) and upstream (◆) direction, and for NG-PON2-induced noise (dashed lines) for down- (●) and upstream (▲) direction. In both standards, the Raman noise in downstream direction dominates, confirming the original assumption of implementing the QKD channel in upstream direction. Looking at this particular case, GPON-induced noise for $C_{US}$ vanishes below the dark counts at 1430 nm (◆), while that due to $C_{DS}$ (◇) is found to be negligible for wavelength longer than its S-bad seed. Thus, $\lambda_{QKD}$ can be safely allocated at the C-band. The NG-PON2 induced Raman noise in upstream direction shows a blanked O-band spectrum due to $C_{DS} + C_{FH}$ (△). $C_{US}$ dominates the total noise and leads to Raman contribution that already reaches the O-band (▲). This tail renders the use of narrowband filters important for the NG-PON2 scenario.

The registered NG-PON2 induced Raman noise counts for various feeder+drop fiber lengths ($L$) and split ratios are shown in Fig. 3a and 3b for detection using the DPS receiver at $\lambda_{QKD}$ = 1310 nm with LAN-WDM and FBG narrowband filter, respectively. With the LAN-WDM filter, Raman noise is pronounced for $L > 5$ km and saturates at 15 km. The FBG provides superior filtering and reduces the Raman-induced counts close to the dark count level. For the original PON setup with 2:16 tree split, the Raman contribution after subtracting the constant dark count rate Δ of the SPAD is 1730 cts/s for the LAN-WDM filter, whereas it falls to ~60 cts/s in the case of the FBG.

The back-to-back QKD results for a dark NG-PON2 scenario are shown in Fig. 3c. The minimum QBER is 1.82% at a loss budget of 12 dB and therefore well below the 5% threshold up to which a secure key can be generated [5]. A raw-key rate of $R$ = 3.4 kb/s and a QBER of 2.9%, well below the QBER limit, are obtained at an elevated budget of 20 dB. SPAD saturation leads to increased QBER at low budgets.

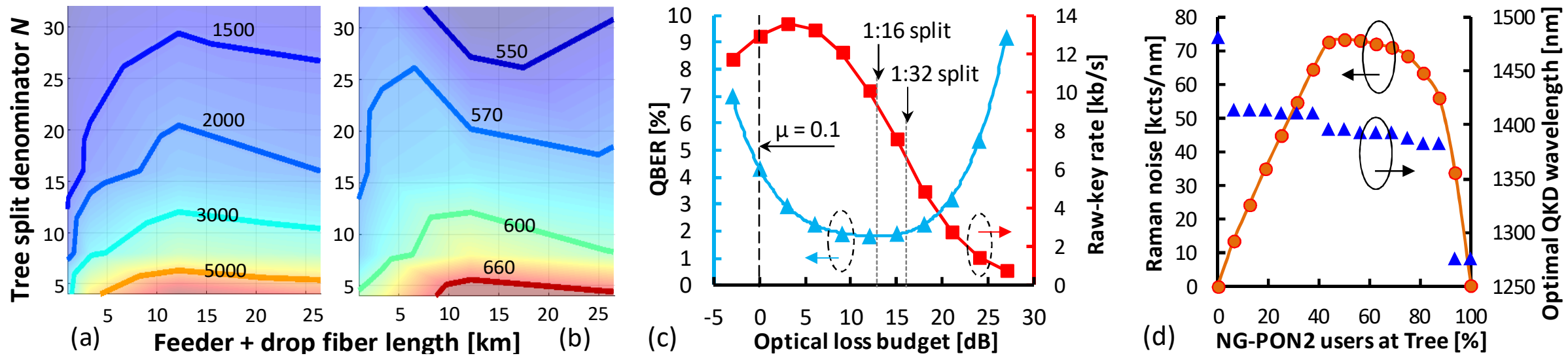

Fig. 3. NG-PON2 induced upstream Raman noise for (a) LAN-WDM and (b) FBG narrowband optical filter at the DPS receiver. (c) Back-to-back QKD performance with in dark NG-PON2 scenario. (d) Optimal $\lambda_{QKD}$ as function of NG-PON2 take-rate at mixed access tree.

The QKD performance for lit PONs are shown in Figs. 4a (GPON) and 4b (NG-PON2). Originating from an average QBER of 3.69% and $R$ = 2.12 kb/s for unloaded GPON (**P**), the loading of any combination of classical channels $C_{US}$ and $C_{DS}$ (**Q**,**R**,**S**) did not show significant influence. Secure-key generation can be accomplished at $3.6\times10^{-7}$ bits/pulse. In case of NG-PON2, the unloaded case (**A**) yields a QBER of 3.16% and $R$ = 3.25 kb/s. The addition of $C_{DS} + C_{FH}$ causes a small QBER penalty of 0.26% with LAN-WDM filter (**B**). However, the impact of $C_{US}$ raises the QBER up to 7.55% (**C**). The use of the FBG alleviates this high Raman noise so that a QBER of ~3.4% at $R$ ~2.5 kb/s remains relatively constant despite the loading of classical channels (**D**,**E**,**F**,**G**). The difference in QBER between DML-based DPS transmitter and an ideal $LiNbO_3$ based phase modulator was 0.1%.

Finally, the case of mixed access traffic is discussed in Fig. 3d for subscription of NG-PON2 users in an GPON-populated ODN. The optimal $\lambda_{QKD}$ assignment is determined by the joint Raman contribution of both access standards. With increasing adoption of NG-PON2 and contamination of the originally blanked C-band, $\lambda_{QKD}$ shifts from the S- to the E-band. This shift is accompanied by a strong increase in Raman noise to >1 kcts/nm due to the closer allocation to GPON sources and new NG-PON2 users. Considering the much lower Raman noise magnitude when having only one of the two access standards present at the tree, mixed traffic is rendered as very detrimental.

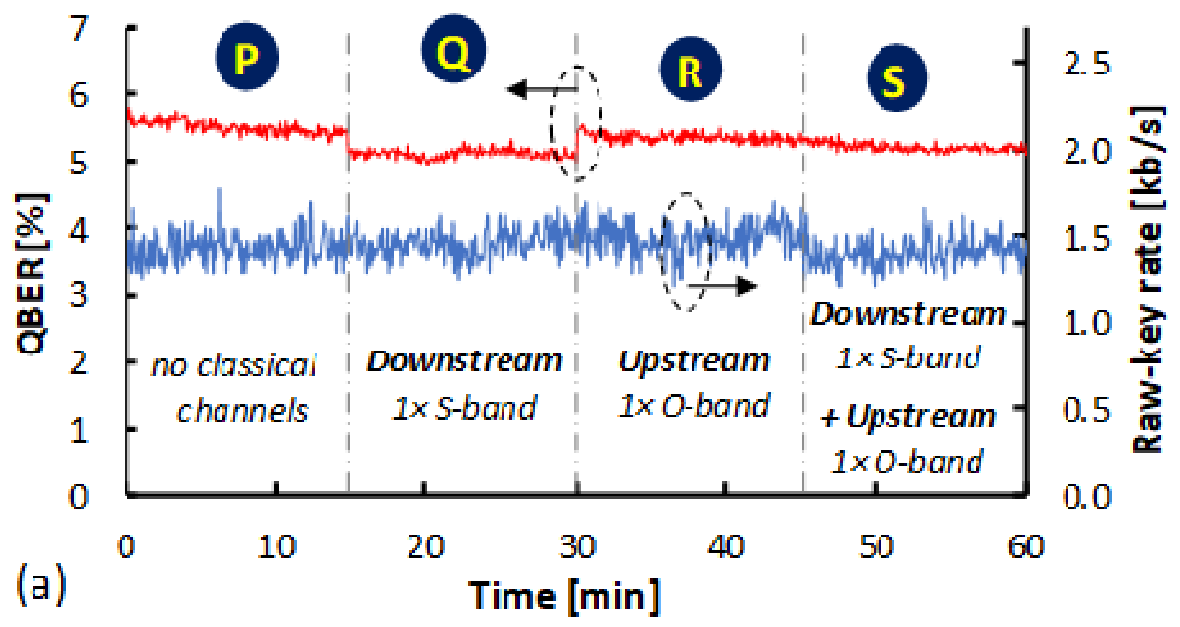

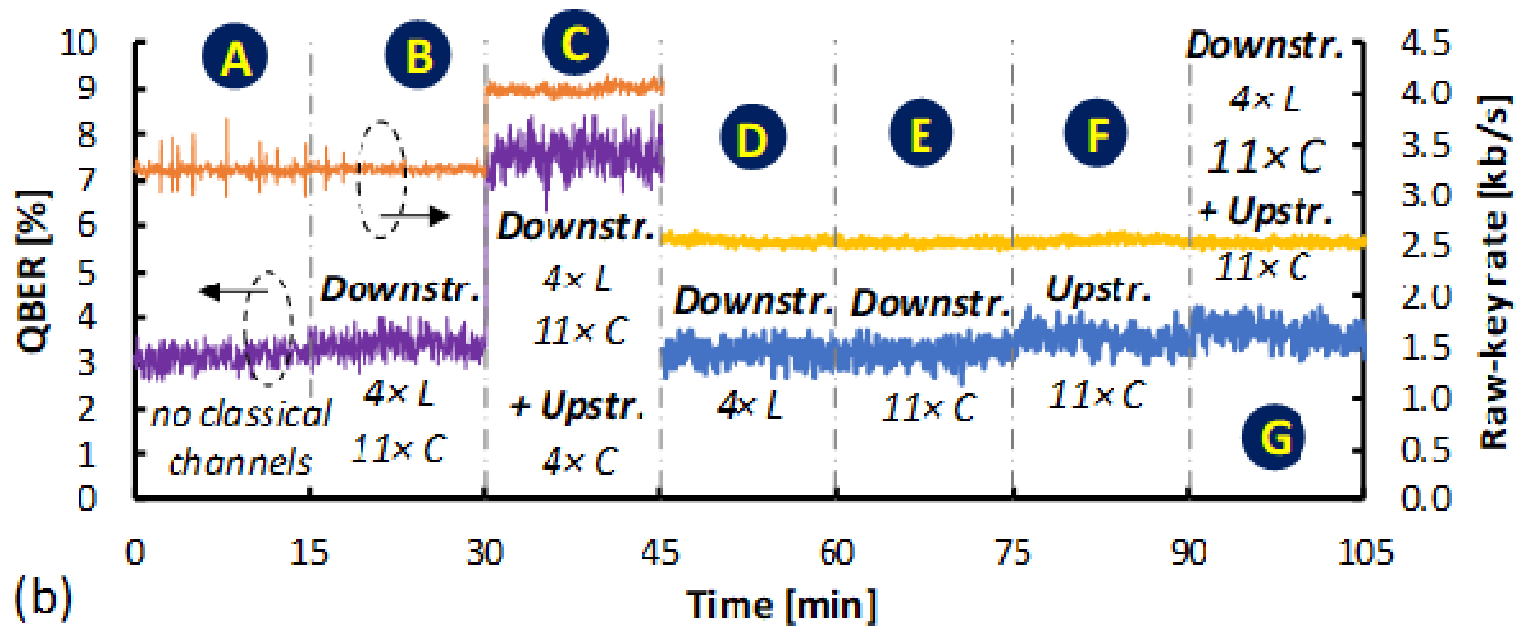

Fig. 4. (a) QKD performance in lit GPON and (b) lit NG-PON2 setting with LAN-WDM and FBG reception filter.

## 5. Conclusions

We have experimentally studied the practical integration of QKD in both, GPON and NG-PON2 access networks with classical channel in the O-, S-, C- and L-band. Using a low-cost DML-based DPS-QKD transmitter, a QBER of 1.82% and a raw-key rate of 10.1 kb/s are obtained at a loss budget of 12 dB. A secure-key rate of 0.51 kb/s is yielded for a 13.5-km reach, 2:16-split PON. Raman noise can be effectively suppressed by means of narrowband filtering. For NG-PON2 with 19 classical channels, the QBER penalty can be kept as low as 0.52%.

## 6. Acknowledgement

This work has received funding from the EU Horizon-2020 R&I programme (grant agreement No 820474).